\documentclass[prb, twocolumn,amsfonts, amssymb, amsmath]{revtex4-1}
\pdfoutput=1
\usepackage{graphicx}
\usepackage{color}
\usepackage{mathtools}
\usepackage{float}

\newcommand{\F}{\op{F}_{\sigma}}  
\newcommand{\M}{\op{M}_{\sigma}}

\newcommand*{\adj}[1]{{#1}^\dagger}  %operator adjoint notation
\newcommand*{\op}[1]{\mathrm{\mathbf{#1}}}  %for formatting operators

\newcommand*{\ket}[1]{\left|#1\right\rangle}
\newcommand*{\expect}[1]{\left\langle#1\right\rangle}
\newcommand*{\bra}[1]{\left\langle#1\right|}
\newcommand*{\braket}[2]{\left\langle#1\middle|#2\right\rangle}
\newcommand*{\ketbra}[3][{}]{\left\vert#2\middle\rangle\raisebox{-0.25em}[0pt][0pt]{\ensuremath{\scriptscriptstyle #1}}\middle\langle#3\right\vert}

\newcommand*{\norm}[1]{\left|#1\right|}

\newcommand*{\Erfc}{\textnormal{Erfc}}

\global\long\def\S{\mathcal{S}}
\global\long\def\A{\mathcal{A}}
\global\long\def\O{\mathcal{O}}
\global\long\def\MI{I\left(\S:\A\right)}

\DeclareMathOperator{\tr}{tr}

\begin{document}
\title{Measurement-induced decoherence and information in double-slit interference}
\author{Joshua Kincaid}
\email{kincajos@oregonstate.edu}
\affiliation{Department of Physics, Oregon State University, Corvallis, OR 97331}

\author{Kyle McLelland}
\affiliation{Department of Physics, Oregon State University, Corvallis, OR 97331}
\author{Michael Zwolak}
\email{mpz@nist.gov}

\affiliation{Center for Nanoscale Science and Technology,
             National Institute of Standards and Technology,
             Gaithersburg, MD 20899}

\affiliation{Department of Physics, Oregon State University, Corvallis, OR 97331}

\begin{abstract}
The double slit experiment provides a classic example of both interference and the effect of observation in quantum physics. When particles are sent individually through a pair of slits, a wave-like interference pattern develops, but no such interference is found when one observes which ``path'' the particles take. We present a model of interference, dephasing, and measurement-induced decoherence in a one-dimensional version of the double-slit experiment. Using this model, we demonstrate how the loss of interference in the system is correlated with the information gain by the measuring apparatus/observer. In doing so, we give a modern account of measurement in this paradigmatic example of quantum physics that is accessible to students taking quantum mechanics at the graduate or senior undergraduate levels.
\end{abstract}

\maketitle

\section{Introduction}
There has been continual interest in the double slit experiment for over two centuries, from its original incarnation for light\cite{Young1802} to its reincarnation after the advent of quantum mechanics.\cite{Joensson1961,Tonomura1989} More generally, interference in quantum systems continues to be an active area of research, as experimentalists endeavor to coax ever larger objects to interfere,\cite{Arndt1999,Eibenberger2013,Arndt2014} to see interference in ion-trap and cold atom systems,\cite{Beugnon2006, Maunz2007} and to understand interference in exotic situations such as Bose-Einstein condensates\cite{Andrews1997} or with topological defects.\cite{Dziarmaga2012}

\begin{figure}[H]
  \begin{center}
    \includegraphics[width=0.9\columnwidth]{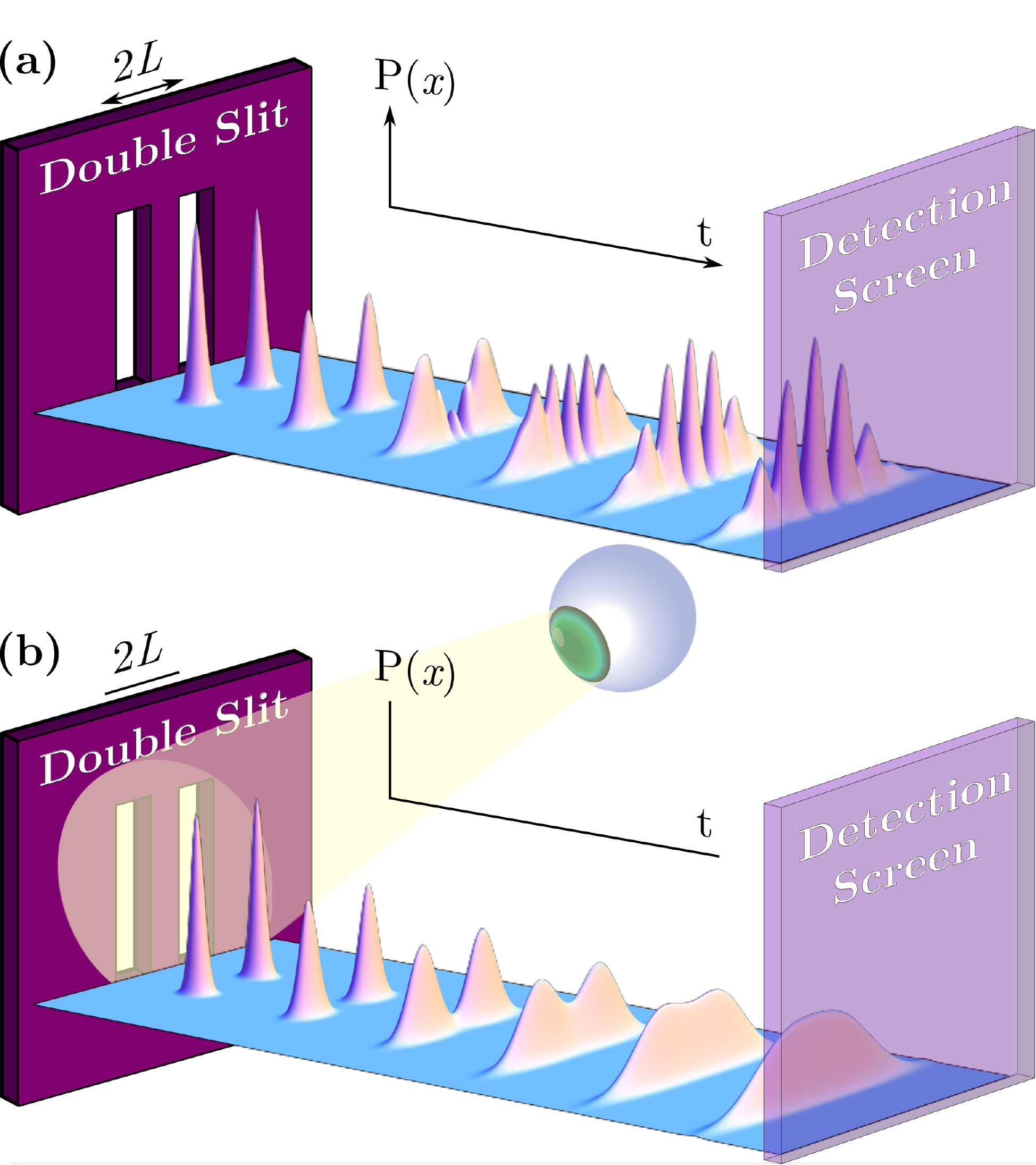}
    \caption{Schematic representation of the one dimensional version of the double slit experiment. In this setup, a coherent superposition of two Gaussian wavepackets of width $\Delta$ emerge from the two slits separated by a distance $2L$. (a) When no measurement is made and coherence is otherwise preserved, the probability density $P(x)$ shows that as time progresses the two packets begin to interfere, ultimately resulting in a well-defined interference pattern. (b) In the case of a perfect measurement, each particle takes either the left or the right path. In this case $P(x)$ observed at the detection screen will be an incoherent sum of the two spreading Gaussian wavepackets, i.e., no interference will be present.}
    \label{fig:schematic}
  \end{center}
\end{figure}

The double-slit experiment is the paradigmatic example of quantum mechanical ``weirdness.'' According to Feynman, it contains ``the only mystery [of quantum mechanics]'' where the particles---whatever they may be, electrons, photons, etc.---behave ``sometimes like a particle and sometimes like a wave.''\cite{Feynman1963c} In the conceptually simplest set-up, a stream of particles is incident on a barrier in which two slits have been made to allow passage. On the far side of the barrier is a detection screen, which is ultimately examined to determine the outcome of the experiment.  As is well known, the detection screen will reveal an interference pattern if no attempt---by an observer or otherwise, e.g., the environment---was made to determine which path was taken by the particles, even when the particles are sufficiently separated in time that only one is ever in the region containing the barrier and the screen. The interference pattern disappears if a detection apparatus is placed to determine which path is taken by each particle. In other words, acquiring a particle's ``which-path'' information prevents it from exhibiting interference. A schematic is shown in Fig.~\ref{fig:schematic}.

If however, one is willing to forgo \emph{perfect} determination of the path, partial interference can be preserved. This relationship has previously been demonstrated in theoretical analyses and proposed experimental realizations of the double-slit experiment for light,\cite{Wootters1979, Raymer1992} and a general treatment of imperfect two-state discrimination from basic quantum mechanical principles can be found in textbooks.\cite{Note1} In short, when the apparatus/observer acquires information about the system the interference pattern disappears.

This deep relationship between interference and information can be understood in the context of decoherence and entanglement, and plays a significant role in understanding the quantum-to-classical transition.\cite{Ollivier2004,Ollivier2005,Blume-Kohout2006,Zurek2009,Zwolak2013,Zwolak2014,Zurek2014} While there exist a multitude of papers\cite{Zurek1981, Zurek1982, Zurek1991, Albrecht1992, Tegmark1994, Schumacher1996, Zurek2003} and books\cite{Joos2003, Nielsen2010, Schlosshauer2010} on the more general subjects of quantum information and decoherence, the explicit application of these ideas to specific physical systems at a level suitable for students is lacking. To that end, we examine a model of double-slit interference in the presence of measurement. The model allows the measurement precision to be tuned, and thus to examine the interplay between path information gained and the loss of interference: When one distinguishes between a particle at the left and right slits, then the interference is destroyed; when no information is gained, then interference is manifest. This is done in an example that is approachable by senior undergraduate and first year graduate students, and thus should help make the core concepts of interference, measurement, distinguishability, decoherence, and dephasing more concrete in the classroom.

In order to keep the discussion concise, we assume that the reader is familiar with certain mathematical and conceptual tools not necessarily presented in introductory courses, the foremost of which are density operators,\cite{Note2} the evaluation of Gaussian integrals,\cite{Shankar1994} and the partial trace.\cite{Note4} We also provide only a brief introduction to quantum entropy and mutual information.\cite{Note5} Students who have not had prior exposure to these concepts may require a brief overview before pursuing a detailed understanding of the model. It is also convenient to work with dimensionless parameters, but we wish to retain the traditional symbols for readability. To that end, we fix a length scale \(\Delta\) (the slit width) and denote physical quantities that carry a dimension with an overbar. The unbarred version is then the natural dimensionless parameter determined by \(\Delta\) and physical constants. Hence we have dimensionless position \(x = \bar{x}/\Delta\), time \(t = \hbar\bar{t}/m\Delta^2\), momentum \(p = \bar{p}\Delta/\hbar\), and so on. Similarly, we use a dimensionless Hamiltonian, \(\op{H} = (m\Delta^2/\hbar^2)\op{\bar{H}}\), momentum operator, \(\op{P} = (\Delta/\hbar)\op{\bar{P}}\), and position operator, \(\op{X} = \op{\bar{X}}/\Delta\).

\section{Interference without measurement\label{sec:Free_Interference}}
We first consider the case where no measurement is attempted. The prototypical version of the double slit experiment is to have particles impinge on a barrier one by one. The barrier has two slits that let particles through, where they then continue to travel until striking the detection screen. The latter will reveal the interference pattern---or lack thereof---that emerges after many repetitions of the experiment. We will consider a simplified version of this scenario where, as indicated in Fig.~\ref{fig:schematic}, there is just one spatial dimension and the evolution starts after the single particle exits the double slit.

When the particle---the system \(\S\)---exits the slits in a superposition of two Gaussian states,\cite{Note6} i.e., its state \(\ket{\Psi}_\S\) is given by the wavefunction
\begin{equation}\label{eq:wave}
\braket{x}{\Psi}_\S = \Psi(x) = A\left[{e}^{-(x + L)^2/2} + {e}^{-(x - L)^2/2}\right],
\end{equation}
the two Gaussian components will begin to spread. Here, as throughout, \(x\) and \(L\) are dimensionless parameters that correspond to position and slit-spacing ($2L$), respectively, and which depend implicitly on the slit-width. The normalization is
\begin{equation}
  A = \sqrt{\frac{1}{2\sqrt{\pi}\left[1 + \exp{\left(-L^2\right)}\right]}}.
\end{equation}

Note that time also represents the role of a second spatial dimension---the one in which the source, barrier, and screen are separated. As time moves on, one can imagine the particle moving from the barrier to the detection screen. More formally, one could include the additional spatial dimensions and integrate them out, as they do not play an important role.

\begin{figure}[H]
  \begin{center}
    \includegraphics[width=0.9\columnwidth]{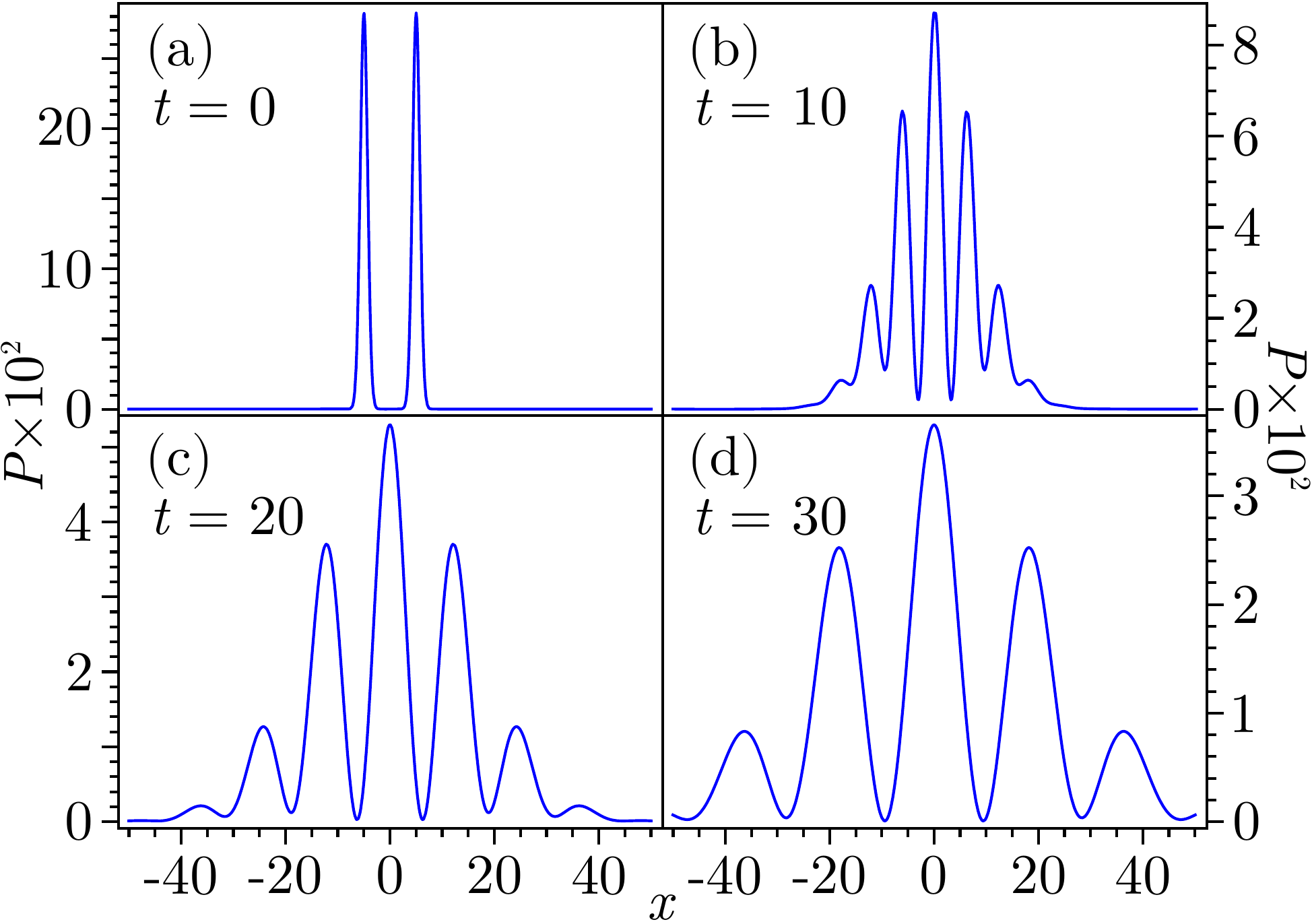}
    \caption{Probability density of a free particle with \(L = 5\) at different times after passing through two slits. Note that this corresponds to the time of flight from the slit to the detection screen. (a) Initially, the two packets are well separated. (b) As they spread, they will start to interfere. (c) Eventually, a well-defined interference pattern develops, which (d) begins to spread out.}
    \label{fig:twoslits}
  \end{center}
\end{figure}

Making use of the free-particle Hamiltonian \(\op{H} = \op{P}^2 / 2\), we find the time-evolved state by integrating
\begin{equation}
  \Psi(x,t) = \frac{1}{2\pi}\int_{-\infty}^{\infty}{\int_{-\infty}^{\infty}{{e}^{-ip^2t/2}{e}^{{i} p(x - x')}\Psi(x') \, dx'} \, dp}.
\end{equation}
 Using the initial wavefunction and evaluating the resulting Gaussian integral, one finds that the time-dependent wavefunction is given by
\begin{align}
  \Psi\left(x,t\right) &= \frac{A}{\sqrt{1 + {i} t}}\left\{\exp{\left[-\frac{\left(x + L\right)^2}{2(1 + {i} t)}\right]}\right.\nonumber\\
  &\left.+ \exp{\left[-\frac{\left(x - L\right)^2}{2(1 + {i} t)}\right]}\right\}.
\end{align}
The associated probability density is then
\begin{align}\label{eq:freefly}
  P\left(x,t\right) &= \Gamma^{xt}\left[\cosh{\left(\frac{2xL}{1 + t^2}\right)} + \cos{\left(\frac{2txL}{1 + t^2}\right)}\right],
\end{align}
where the factor
\begin{equation}
\Gamma^{xt} = \frac{2 A^2}{\sqrt{1 + t^2}}\exp{\left(\frac{-x^2 - L^2}{1 + t^2}\right)}
\end{equation}
has been introduced for readability. The first term in Eq. (\ref{eq:freefly}), the one with the hyperbolic cosine, is just a sum of two Gaussian wavepackets, which represent particles coming from the left or right slit, respectively. The second term, the one with the cosine, describes the interference between these two sources. Figure~\ref{fig:twoslits} shows the probability density, Eq. (\ref{eq:freefly}), for various times---i.e., the separation between the slits and detection screen---for a slit spacing \(2L\gg1\).

\section{The effect of measurement\label{ssec:MMM}}
We now want to consider how measurement affects the appearance of interference. Specifically, we are interested in how the interference pattern is lost as the amount of information gained increases. In an introductory quantum course, one would discuss, for example, double-slit interference of electrons and measurement via photons. In this case, the observation is performed by measuring scattered light, which, with a suitably short wavelength, distinguishes the position of the electron. As the wavelength is increased, the scattered light no longer is imprinted with the left/right position of the electron. Calculating this scattering process, however, requires a large amount of background material, unsuitable for introductory courses.

Instead, we consider an idealized measurement process that nevertheless admits variable precision ranging continuously from perfect measurement to no measurement at all. Following the idea of a von Neumann chain,\cite{Neumann1932} this process will make use of an auxiliary quantum system, the apparatus \(\A\). When the apparatus, in an appropriately initialized state \(\ket{0}_\A\), interacts with a particle in the state \(\ket{\psi^L}_\S\) with wavefunction \(\psi^L(x)\) that is completely localized around the left slit (where we use the less strict condition, \(\psi^L(x) = 0\) for \(x>0\)), a perfect measurement would bring the composite state \(\ket{\psi^L, 0}_{\S\A}\) to \(\ket{\psi^L, L}_{\S\A}\). Similarly, when the apparatus interacts with a particle in a state \(\ket{\psi^R}_\S\) with a wavefunction \(\psi^R(x)\) that is completely localized around the right slit, a perfect measurement would bring the state \(\ket{\psi^R, 0}_{\S\A}\) to \(\ket{\psi^R, R}_{\S\A}\). This process transfers information about the particle's state into \(\A\), encoding the outcome of the left/right-measurement in a subspace of dimension two spanned by the basis states \(\ket{L}_\A\) and \(\ket{R}_\A\). This left/right information is accessible to observers who can ``read'' the apparatus state. If one has a limited resolution measurement, or wavefunctions \(\psi^L(x)\) and \(\psi^R(x)\) that have overlap, then this information transfer cannot be perfect.

\subsection{The measurement interaction\label{ssec:Int}}
In general, after a measuring apparatus \(\A\) interacts with a system \(\S\), observers can infer the state of the system by interacting with (and amplifying information from) the apparatus through the standard measurement process; i.e., by measuring a non-degenerate observable of \(\A\) corresponding to the possible  measurement outcomes. Of course, such a subsequent measurement could be treated similarly, requiring yet another measuring apparatus, and so on, leading one ultimately to the von Neumann chain. We are here concerned only with the first step in such a chain, considering only the interaction between \(\S\) and \(\A\). Later on, we briefly discuss the observer as an additional link in the von Neumann chain.

\begin{figure}[H]
  \begin{center}
    \includegraphics[width=0.9\columnwidth]{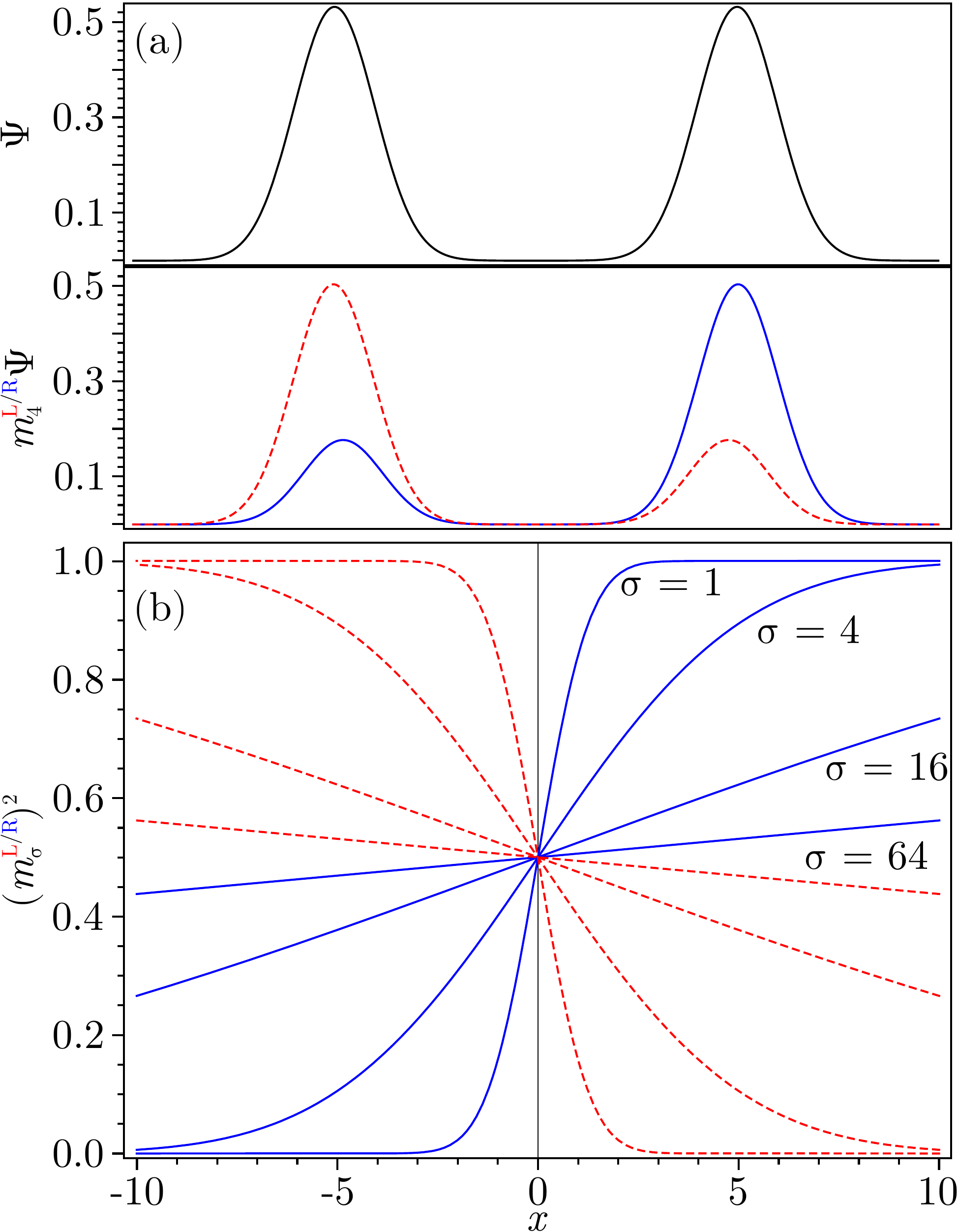}
    \caption{The effect of measurement. (a) The initial wavefunction \(\Psi\) with peaks at \(L = \pm5\) and the conditional states of the post-measurement wavefunction with \(\sigma = 4\), \(\Psi^L = m^L_4\Psi\) (dashed) and \(\Psi^R = m^R_4\Psi\) (solid). (b) The measurement functions squared, \((m^L_\sigma)^2\) (dashed) and \((m^R_\sigma)^2\) (solid), versus position. As \(\sigma\) increases and the measurement becomes less precise, they tend toward a common constant value; in the opposite limit, they become complementary step-functions.}
    \label{fig:mfunc}
  \end{center}
\end{figure}

In our case, the relevant (non-degenerate) eigenstates of \(\A\) are \(\ket{L}_\A\) and \(\ket{R}_\A\). We assume that the apparatus and system interact immediately after the particle passes through the slit, so that the particle wavefunction, Eq. (\ref{eq:wave}), does not have time to evolve on its own before the measurement is made. As usual, the interaction between the apparatus and the system results in a unitary transformation of the joint state. Specifically, in keeping with the above discussion, we require that during the measurement process the joint state, initially \(\ket{\Psi,0}_{\S \A}\), evolves as
\begin{equation}\label{eq:unitarymeasure}
  \ket{\Psi,0}_{\S \A} \mapsto \ket{\M^L\Psi, L}_{\S \A} + \ket{\M^R\Psi, R}_{\S \A} \equiv \ket{\Phi}_{\S\A}
\end{equation}
during the interaction (here, as elsewhere, we use \(\Psi\) for our specific system state in distinction to the \(\psi\) used for generic system states in the introduction to Sec.~\ref{ssec:MMM}). When the apparatus registers ``$L$'', the system will be in a state $\ket{\Psi^L}\propto\M^L \ket{\Psi}_{\S}$ that is localized--- to a precision $\sigma$---around the left slit due to the act of measurement itself. The conditional state $\ket{\Psi^L}$ depends on both the initial system state and the measurement operator (similarly for $\ket{\Psi^R}\propto\M^R \ket{\Psi}_{\S}$). The initial state and the states \(\Psi^L\) and \(\Psi^R\) resulting from such an interaction are shown in Fig. \ref{fig:mfunc} (a) using an explicit form of the measurement operator to be derived later in Eq.~(\ref{eq:measuredensity}). The right-hand side of Eq. (\ref{eq:unitarymeasure}) cannot, in general, be written as a simple product of system and apparatus states. Thus, the interaction has caused the two to become \emph{entangled} (except, of course, in the limiting case of no discrimination).

The operators \(\M^{L/R}\) appearing in Eq. (\ref{eq:unitarymeasure}) are called ``measurement operators'' and are written inside the ket in order to make clear the fact that they act only on the system \(\S\). As noted, these operators determine the state of the particle after the measurement. Since the left/right measurement distinguishes the position of the particle, it suffices to take \(\M^{L/R}\) diagonal in the position basis, giving rise to a pair of ``measurement functions''
\begin{align}\label{eq:mfuncdef}
\M^L\ket{x} &= m_\sigma^L(x)\ket{x},\nonumber\\
\M^R\ket{x} &= m_\sigma^R(x)\ket{x}.
\end{align}
With this choice, the action of \(\M^{L/R}\) on the system wavefunction is purely multiplicative, \(\bra{x}\M^{L/R}\ket{\Psi} = m^{L/R}_\sigma(x)\Psi(x)\). Since the apparatus states \(\ket{L}_\A\) and \(\ket{R}_\A\) are orthogonal, the requirement that Eq. (\ref{eq:unitarymeasure}) constitutes a \emph{unitary} transformation will be satisfied whenever
\begin{equation}\label{eq:measure_complete}
\adj{\M^L}\M^L + \adj{\M^R}\M^R = \op{I},
\end{equation}
which, from Eq. (\ref{eq:mfuncdef}), is equivalent to
\begin{equation}\label{eq:mfuncs}
  \norm{m^L_\sigma(x)}^{2} + \norm{m^R_\sigma(x)}^{2} = 1.
\end{equation}
Equation (\ref{eq:unitarymeasure}) can then be extended to a unitary transformation defined on the whole joint Hilbert space. We note that, while Eq. (\ref{eq:unitarymeasure}) does not uniquely determine this unitary transformation on the whole space, it suffices as a description of the interaction when the apparatus is initialized to \(\ket{0}_\A\).

While the composite state evolves unitarily, we are also interested in the states of the system and apparatus separately. In particular, we are interested in whether the system state exhibits interference and how the apparatus state encodes information about the system. For this reason we will examine the density operator \(\rho_{\S\A} = \ketbra[\S\A]{\Phi}{\Phi}\) and the reduced states of the system and apparatus. Taking the partial trace\cite{Note7} gives
\begin{align}\label{eq:poststate}
  \rho_\S &= \tr_\A{\rho_{\S\A}} = \prescript{}{\A}{\bra{L}}\rho_{\S\A}\ket{L}_\A + \prescript{}{\A}{\bra{R}}\rho_{\S\A}\ket{R}_\A\nonumber\\
  &= \M^L\ketbra[\S]{\Psi}{\Psi}\adj{\M^L} + \M^R\ketbra[\S]{\Psi}{\Psi}\adj{\M^R}\nonumber\\
  &\equiv \frac{1}{2}\ketbra[\S]{\Psi^L}{\Psi^L} + \frac{1}{2}\ketbra[\S]{\Psi^R}{\Psi^R},
\end{align}
which is a mixture of the states corresponding to distinct detection outcomes. Note that the degree of overlap between the states \(\ket{\Psi^{L}}\) and \(\ket{\Psi^{R}}\) depends on the measurement precision; they are orthogonal when the measurement perfectly distinguishes left from right (\(\sigma = 0\)) but identical in the opposite limit of \(\sigma\to\infty\) (in which case \(\M^L = \M^R = \op{I}/\sqrt{2}\) and no actual measurement is made). Hence, except in this latter case, the system transitions from a coherent superposition or pure state (i.e., one representable by a ket) to a mixed state (one which cannot be represented by a ket): it has been \emph{decohered},\cite{Zurek2003} to an extent that depends on the measurement precision, through its interaction with the apparatus.

The partial trace over the system can be evaluated in the position basis with the help of Eq. (\ref{eq:mfuncdef}), leading to the apparatus state
\begin{equation}\label{eq:postapp}
  \rho_{\A} = \begin{pmatrix}\int{\norm{m^L_\sigma}^2\norm{\Psi}^2 d \, x}&\int{m^R_\sigma m^L_\sigma\norm{\Psi}^2 d \, x}\\\int{m^L_\sigma m^R_\sigma\norm{\Psi}^2 d \, x}&\int{\norm{m^R_\sigma}^2\norm{\Psi}^2d \, x}\end{pmatrix},
\end{equation}
when represented in the \(\{\ket{L}_\A, \ket{R}_\A\}\) subspace.

Further evaluation requires a definite choice of \(m^{L/R}_\sigma(x)\). While there are many possibilities, a physically meaningful choice can be made by considering first a continuous position measurement,\cite{Caves1987,Dowker1992} out of which we can build a coarse-grained, binary measurement. To that end, consider the position-indexed, commuting set of operators \(\F(x')\), defined by
%\begin{equation}
%  \F(x) = \frac{1}{\sqrt{2\pi\sigma^2}}\int_{-\infty}^{\infty}{\exp{\left(\frac{-(x - x')^2}{2\sigma^{2}}\right)}\ketbra{x'}{x'} \, dx'},
%\end{equation}
\begin{equation}
  \F(x')\ket{x} = \frac{1}{\sigma\sqrt{2\pi}}\exp{\left(\frac{-(x - x')^2}{2\sigma^{2}}\right)}\ket{x},
\end{equation}
which represent a smooth analog of the projection operator \(\ketbra{x'}{x'}\) (to which \(\F(x')\) tends as \(\sigma\to0\)).

By integrating separately over the positive and negative domains of \(\F\), we arrive at a pair of coarse-grained operators acting on the position basis as
\begin{align}
  \F^{L}\ket{x} &= \int^0_{-\infty}{\F(x')\ket{x} \, dx'}\nonumber\\
  &= \left(\frac{1}{\sqrt{\pi}}\int_{x/\sqrt{2\sigma^2}}^{\infty}{{e}^{-u^2} \, du}\right)\ket{x},\nonumber\\
  & = \frac{1}{2}\Erfc{\left(\frac{x}{\sigma\sqrt{2}}\right)}\ket{x},\label{eq:Erfc}\\
  \F^{R}\ket{x} &= \int^{\infty}_0{\F(x')\ket{x} \, dx'} = \frac{1}{2}\Erfc{\left(\frac{-x}{\sigma\sqrt{2}}\right)}\ket{x},
\end{align}
where we made use of the complementary error function \(\Erfc\). These operators correspond to left and right positions with precision $\sigma$.

Comparing to Eqs. (\ref{eq:measure_complete}) and (\ref{eq:mfuncs}), we see that taking
\begin{equation}\label{eq:measuredensity}
   m^L_\sigma(-x) = m^R_\sigma(x) \equiv m_\sigma(x) = \left[\frac{1}{2}\Erfc{\left(\frac{-x}{\sigma\sqrt{2}}\right)}\right]^{1/2}
\end{equation}
yields measurement operators satisfying \(\adj{\M^{L/R}}\M^{L/R} = \F^{L/R}\). One can check directly that the pair of operators \(\F^{L/R}\) satisfy \(\F^L + \F^R = \op{I}\), so we conclude that Eq. (\ref{eq:measuredensity}) provides a physically meaningful function that satisfies our criteria. Figure \ref{fig:mfunc} shows how the function \(m_\sigma^2\) changes as one varies \(\sigma\).

Returning to Eq. (\ref{eq:postapp}) and inserting Eq. (\ref{eq:measuredensity}), one finds that the diagonal terms evaluate to  Gaussian integrals, which can be computed exactly. The off-diagonal terms contain the product \(m_\sigma^L m_\sigma^R\), which does not result in a simple closed-form expression (although it is easily evaluated numerically for specific values of \(\sigma\)). In order to obtain an analytic expression for arbitrary \(\sigma\), some approximation will be necessary. To that end, recall that \(\Psi\) is a superposition of two Gaussians centered at \(L\) and \(-L\), respectively. For \(L \gg 1\), the value of \(m_\sigma\) changes little over the regions in which \(\Psi\) is non-negligible. This can be seen qualitatively by considering the curves in Fig. \ref{fig:mfunc} or analytically by expanding \(m_\sigma\) in a Taylor series about \(x = \pm L\). For example, expanding around \(x = L\), we find
\begin{align}\label{eq:linear_m}
m_\sigma(x) \approx m_\sigma(L) + \frac{{e}^{-L^2/2\sigma^2}}{\sigma\sqrt{8\pi}m_\sigma(L)}(x - L).
\end{align}
Then for \(\sigma > 0\), \(m_\sigma(L)\) is bounded below by \(1/\sqrt{2}\). Substituting this lower bound for \(m\) and the maximizing the derivative with respect to \(\sigma\) shows that, for fixed \(L\),
\begin{equation}
\frac{{e}^{-L^2/2\sigma^2}}{\sigma\sqrt{8\pi}m_\sigma(L)} \leq \frac{1}{2L\sqrt{\pi{e}}} < \frac{1}{5L}.
\end{equation}
Hence, the linear coefficient here is certainly smaller than \(0.2/L\) for all \(\sigma\), i.e., over the width of the Gaussian ($1$ in the dimensionless units employed here).  The first-order change to $m_\sigma(x)$ near \(L\) is at most \(0.2/L\) (this is a worst case estimate, and for $L=5$ gives a bound of \(0.04\)). Higher order terms are likewise suppressed. Similarly, the first-order term near \(x = -L\) is bounded by \(0.4/L\).\cite{Note8} We therefore consider \(m_\sigma(x){e}^{-(x \pm L)^2/2} \approx m_\sigma(\mp L){e}^{-(x \pm L)^2/2}\) and \(m_\sigma(-x){e}^{-(x \pm L)^2/2} \approx m_\sigma(\pm L){e}^{-(x \pm L)^2/2}\); i.e., the functions \(m_\sigma(x)\) are approximated, but not the Gaussian envelopes. This results in the approximations
\begin{align}\label{eq:approxfuncL}
  \Psi^L(x) &= \braket{x}{\Psi^L} = \sqrt{2}m_\sigma^L(x)\Psi(x) \nonumber\\
  &= A\sqrt{2}[m_\sigma(-x){e}^{-(x + L)^2/2} + m_\sigma(-x){e}^{-(x - L)^2/2}]\nonumber\\
  &\approx B[m_\sigma(L){e}^{-(x + L)^2/2} + m_\sigma(-L){e}^{-(x - L)^2/2}],
\end{align}
and
\begin{align}\label{eq:approxfuncR}
  \Psi^R(x) &= \braket{x}{\Psi^{R}} = \sqrt{2}m_\sigma^R(x)\Psi(x) \nonumber\\
  &= A\sqrt{2}[m_\sigma(x){e}^{-(x + L)^2/2} + m_\sigma(x){e}^{-(x - L)^2/2}]\nonumber\\
  &\approx B[m_\sigma(-L){e}^{-(x + L)^2/2} + m_\sigma(L){e}^{-(x - L)^2/2}].
\end{align}
Note that in both $\Psi^L(x)$ and $\Psi^R(x)$, each term in the superposition is approximated separately. 

A straightforward integration shows that the  normalization constant of the approximate states should be
\begin{equation}\label{eq:newnorm}
B^{-2} = \sqrt{\pi}\left(1 + \beta_\sigma{e}^{-L^2}\right) \approx \sqrt{\pi},
\end{equation}
where the approximation is for \(L \gg 1\) (i.e., \({e}^{-L^2} \ll 1\)), and
\begin{equation}\label{eq:beta}
  \beta_\sigma = 2m_\sigma(-L)m_\sigma(L).
\end{equation}
As we will show, \(\beta_\sigma\), which ultimately depends on both \(\sigma\) and \(L\) through the ratio \(\sigma/L\), is the parameter that relates the measurement precision to the visibility of interference fringes and the information acquired by the measurement apparatus. In some sense, one can think of  \( \beta_\sigma \) as the relevant quantification of the overlap between the left and right measurements.

Returning again to Eq. (\ref{eq:postapp}), it is clear that the approximation of Eqs. (\ref{eq:approxfuncL}) and (\ref{eq:approxfuncR}), does not affect the diagonal terms. It does, however, allow us to evaluate the off-diagonal terms as intended, which are now also just Gaussian integrals. Doing so, we find
\begin{equation}
  \op{\rho}_\A = \frac{1}{2}\begin{pmatrix}1&\frac{\beta_\sigma + {e}^{-L^2}}{1 + \beta_\sigma{e}^{-L^2}}\\\frac{\beta_\sigma + {e}^{-L^2}}{1 + \beta_\sigma{e}^{-L^2}}&1\end{pmatrix} \approx \frac{1}{2}\begin{pmatrix}1&\beta_\sigma\\\beta_\sigma&1\end{pmatrix}.
\end{equation}
This has trace \(1\), as expected, and the eigenvalues are
\begin{equation}\label{eq:appeigens}
  \lambda_{\pm} = \frac{1}{2}\left(1 \pm \frac{\beta_\sigma + {e}^{-L^2}}{1 + \beta_\sigma{e}^{-L^2}}\right)\approx\frac{1}{2}(1 \pm \beta_\sigma).
\end{equation}
As in Eq.~(\ref{eq:newnorm}), the approximate expressions are for \(\exp(-L^2) \ll 1\), but we retain a finite \(\sigma/L\) in \( \beta_\sigma \) in order to investigate the full range of measurement precision.

\subsection{Post-measurement evolution}\label{ssec:evo}

It is well known that the act of measuring exactly which path a particle takes in passing through the slits prevents the appearance of interference effects. Having determined the immediate effect of measurement on the particle's state, we must now evaluate the subsequent evolution in order to determine how the interference is affected.

After the measurement, the system evolves as a free particle while the apparatus remains unchanged.  The joint evolution is
\begin{align}
 \ket{\Psi, 0} &\mapsto \ket{\M^L\Psi, L}_{\S \A} + \ket{\M^R\Psi, R}_{\S \A}\nonumber\\
  &\mapsto\frac{1}{\sqrt{2}}\ket{\op{\mathcal{U}}_t\Psi^L, L} + \frac{1}{\sqrt{2}}\ket{\op{\mathcal{U}}_t\Psi^R, R},
\end{align}
with \(\op{\mathcal{U}}_t = {e}^{-{i}\op{P}^2t/2}\) and \(\ket{\Psi^{L/R}}\) given in Eq. (\ref{eq:poststate}). We thus have a system state comprising an equal mixture of the wavefunctions
\begin{align}\label{eq:wavefunction}
  \Psi^{L/R}(x, t) &= \bra{x}{\op{\mathcal{U}}_t}\ket{\Psi^{L/R}}\nonumber\\
  &= \int{{e}^{-\frac{{i}}{2}p^2t}\braket{x}{p}\int{\braket{p}{y}\braket{y}{\Psi^{L/R}} \, dy} \, dp}.
\end{align}
Evaluation of the inner integral can be done with approximations (\ref{eq:approxfuncL}) and (\ref{eq:approxfuncR}) and \(\braket{x}{p} = {e}^{{i} xp}/\sqrt{2\pi}\), which reduces the integrals appearing in Eq. (\ref{eq:wavefunction}) to a sum of Gaussian integrals. These give
\begin{align}
  \Psi^L(x,t) = &\left(\frac{B^2}{1 + {i} t}\right)^{1/2}\left\{m_\sigma(-L)\exp{\left[\frac{-\left(x + L\right)^{2}}{2\left(1 + {i} t\right)}\right]}\right.\notag\\
  &\left.+ m_\sigma(L)\exp{\left[\frac{-\left(x - L\right)^{2}}{2\left(1 + {i} t\right)}\right]}\right\},
\end{align}
and
\begin{align}
  \Psi^R(x,t) = &\left(\frac{B^2}{1 + {i} t}\right)^{1/2}\left\{m_\sigma(L)\exp{\left[\frac{-\left(x + L\right)^{2}}{2\left(1 + {i} t\right)}\right]}\right.\notag\\
  &\left.+ m_\sigma(-L)\exp{\left[\frac{-\left(x - L\right)^{2}}{2\left(1 + {i} t\right)}\right]}\right\}.
\end{align}
Recalling that our particle  is in an equal mixture of these two states, the probability density associated with detecting the particle at position \(x\) is given by
\begin{align}
  P_\sigma(x,t) &= \frac{1}{2}\norm{\Psi^L(x,t)}^{2} + \frac{1}{2}\norm{\Psi^R(x,t)}^{2}\notag\\
  &= \Gamma^{xt}_\sigma\left[\cosh{\left(\frac{2xL}{1 + t^2}\right)} + \beta_\sigma\cos{\left(\frac{2txL}{1 + t^2}\right)}\right],
\end{align}
where we have reintroduced and generalized
\begin{align}
  \Gamma^{xt}_\sigma &= \frac{\exp{\left(\frac{-x^2 - L^2}{1 + t^2}\right)}}{\sqrt{\pi + \pi t^2}\left(1 + \beta_\sigma{e}^{-L^2}\right)}\\
  &\approx \frac{\exp{\left(\frac{-x^2 - L^2}{1 + t^2}\right)}}{\sqrt{\pi}\sqrt{1 + t^2}}\nonumber,
\end{align}
for readability. As for the unmeasured free particle, the first term in the brackets (the hyperbolic cosine) describes the spread of the two incoherent Gaussian wavepackets with time. The second term (with the cosine) gives rise to the interference pattern and is also the same as in Eq. (\ref{eq:freefly}), except for a factor of \(\beta_\sigma\). Hence, we recover the unmeasured case in the limit where the measurement is not at all precise, for $\sigma \to \infty$  (\(\beta_\sigma \to 1\)), and the interference is suppressed for smaller values of \(\sigma\). In the limit of a perfectly precise measurement, \(\sigma \to 0\) (\(\beta_\sigma \to 0\)), the interference term vanishes and we find 
\begin{align}
  P_0(x,t) &= \Gamma^{xt}_0\cosh{\left(\frac{2xL}{1 + t^2}\right)}\nonumber\\
  & \propto \exp{\left[\frac{-(x + L)^2}{1 + t^2}\right]} + \exp{\left[\frac{-(x - L)^2}{1 + t^2}\right]},
\end{align}
which describes an incoherent sum of the particle coming either from the left or right slit as shown in Fig. \ref{fig:schematic}b.

\begin{figure}[H]
  \begin{center}
    \includegraphics[width=0.9\columnwidth]{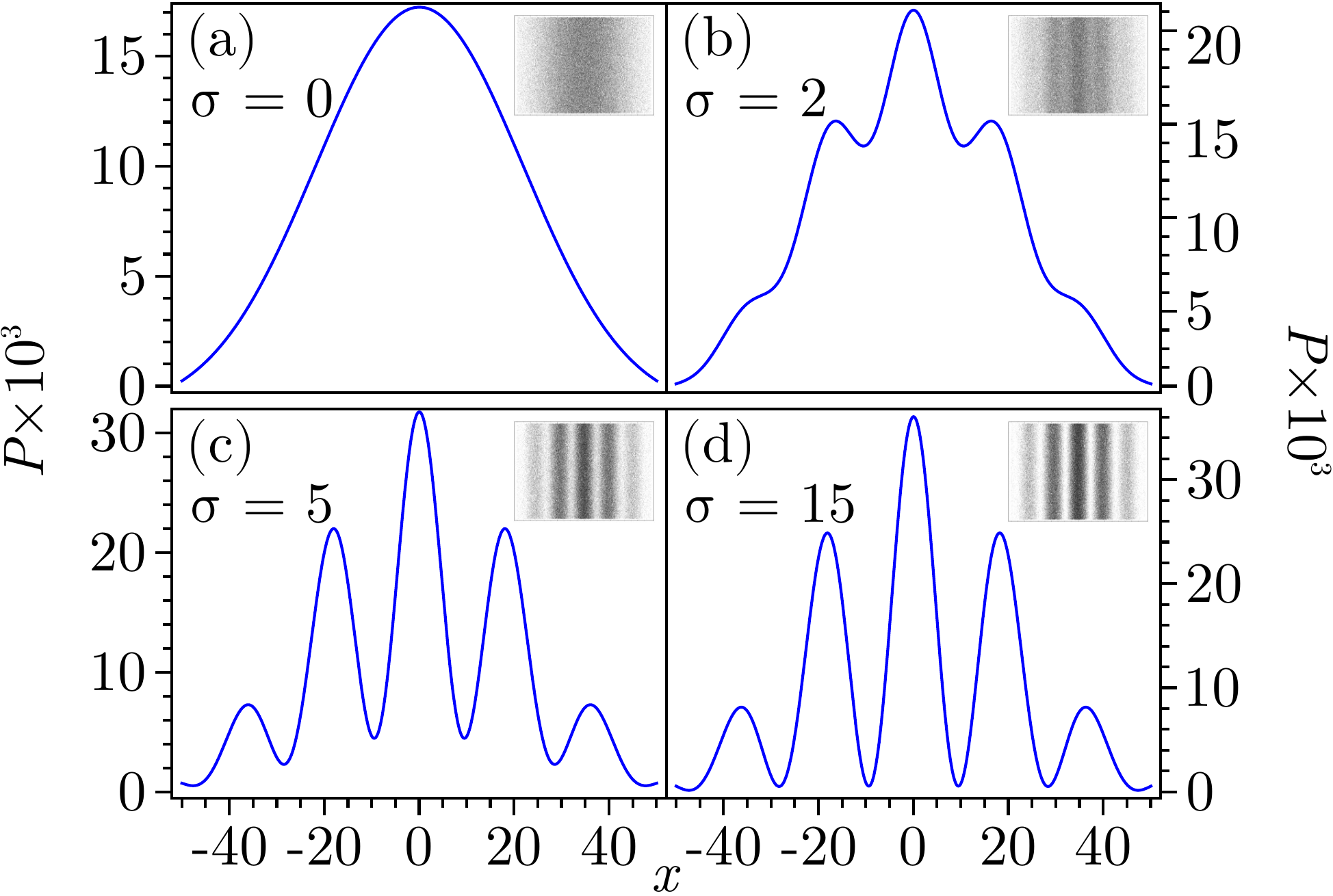}
    \caption{Probability density with \(L = 5\) at \(t = 30\) for various values of \(\sigma\). Insets show simulated detection screens. (a) At \(\sigma = 0\), a perfect measurement has been made; only a single, broad fringe appears. (b) As \(\sigma\) passes \(\sigma^{*} \approx 3L/10\), interference begins to appear. Increasing \(\sigma\) further, (c) the difference between constructively and destructively interfering regions is clearly visible, so that (d) by \(\sigma = 3L\) the interference is nearly total.} 
    \label{fig:measured_particle}
  \end{center}
\end{figure}

Figure \ref{fig:measured_particle} demonstrates how the distribution varies as \(\sigma\) increases.\cite{Note9} Beyond this qualitative demonstration, a quantitative description is provided by the interferometric visibility, which relates the amplitude of a wave to its average value. At sufficiently late times, when a maximum appears at \(x = 0\), this visibility may be expressed as
\begin{equation}\label{eq:vis}
  \mathcal{V} = \frac{P(0, t) - P(x^\ast, t)}{P(0, t) + P(x^\ast, t)},
\end{equation}
where \(x^\ast = \pi(1 + t^2)/2tL\) corresponds to the first minimum of the oscillation term. Evaluating Eq. (\ref{eq:vis}) then yields the expression
\begin{equation}
  \mathcal{V} = \frac{\Gamma_{0t}^\sigma\left[1 + \beta_\sigma\right] - \Gamma_{x^\ast t}^\sigma\left[\cosh{\left(1/t\right)} - \beta_\sigma\right]}{\Gamma_{0t}^\sigma\left[1 + \beta_\sigma\right] + \Gamma_{x^\ast t}^\sigma\left[\cosh{\left(1/t\right)} - \beta_\sigma\right]},
\end{equation}
which, after canceling common factors and taking the limit \(t\to\infty\), reduces to
\begin{equation}\label{eq:visbeta}
  \mathcal{V} = \frac{\left(1 + \beta_\sigma\right) - {e}^{-\pi^2/4L^2}\left(1 - \beta_\sigma\right)}{\left(1 + \beta_\sigma\right) + {e}^{-\pi^2/4L^2}\left(1 - \beta_\sigma\right)}\approx\beta_\sigma.
\end{equation}
Hence, for large \(L\), we have \(\mathcal{V}\approx\beta_\sigma\). Thus, \(\beta_\sigma = 2 m_\sigma (L) m_\sigma (-L) \), which is a quantification of the measurement precision with respect to the slit width.  It further has a  direct physical meaning as the visibility of the post measurement interference fringes.

\section{Information}\label{sec:Information}
We have called the interaction determined in Eq. (\ref{eq:unitarymeasure}) a measurement interaction on the grounds that a subsequent projective measurement of the apparatus alone will allow an observer to infer (or attempt to infer, in the case of an imperfect measurement) the state of the system. This interaction is just a particular example of a positive-operator valued measure (POVM).\cite{Note10} The key idea is that the apparatus acquires information about the state of the system due to this interaction. To make this statement quantitative, we make use of two key ideas from the theory of quantum information: entropy and mutual information.

For any state represented by a density operator \(\op{\rho}\) with eigenvalues \(\{ \lambda_i \}\), the von Neumann entropy is defined by
\begin{equation}
  H(\op{\rho}) = -\tr{\left[\op{\rho}\log_2{\op{\rho}}\right]} = -\sum_i{ \lambda_i \log_2{\lambda_i}},
\end{equation}
in which we take \(0\log{0} = 0\) whenever it arises. In particular, \(H(\op{\rho}) \geq 0\), with equality if and only if the state is pure (i.e., \(\lambda_i = 1\) for one \(i\) and \(\lambda_i = 0\) otherwise). Hence, the entropy is a measure of our state's ``mixedness'' and quantifies our uncertainty about the state of the system. It thus also quantifies the amount of information we gain about the system when a measurement is made.\cite{Note11}

If our system is composed of two subsystems in a state \(\rho_{\S\A}\) and with reduced states \(\op{\rho}_\S\) and \(\op{\rho}_\A\), the quantum mutual information between $\S$ and $\A$ is defined by
\begin{equation}
  I\left(\S:\A\right) = H(\op{\rho}_\S) + H(\op{\rho}_\A) - H(\op{\rho}_{\S\A}).
\end{equation}
This quantifies the amount of information about system \(\S\) that is in \(\A\).

In the case of a measurement implemented by some apparatus as described previously, we have
\begin{align}
  I\left(\S:\A\right) &= H(\op{\rho}_{\S}) + H(\op{\rho}_{\A}) - H(\op{\rho}_{\S\A})\\
  &= H(\op{\rho}_\S) + H(\op{\rho}_\A),
\end{align}
since the joint-state is pure (having evolved unitarily from a pure product state). Moreover, when the joint state is pure, the Schmidt decomposition\cite{Schmidt1908,Ekert1995,Note12} ensures that we can use \(\ket{\Phi}_{\S\A} = \sum_i{a_i\ket{i}_\S\otimes\ket{i}_\A}\), where \(\{\ket{i}_\S\}\) and \(\{\ket{i}_\A\}\) are orthonormal bases for the two subsystems, in writing the joint density matrix \(\rho_{\S\A} = \ketbra[\S\A]{\Phi}{\Phi}\). Taking the partial traces, one can see that the values \(\{a_i\}\) will be the eigenvalues for \emph{both} \(\op{\rho}_\S\) and \(\op{\rho}_\mathcal{\A}\), so that their entropies will be the same. In particular, we have
\begin{equation}\label{eq:infodeco}
  I\left(\S:\A\right) = 2H(\op{\rho}_\A) = 2H(\op{\rho}_\S).
\end{equation}
Again, the entropy \(H(\op{\rho}_\S)\) gives a measure of the mixedness (the degree of decoherence) of our system state after the measurement. This measurement-induced decoherence of the system is associated with information acquisition by the measurement apparatus, which is reflected in this generation of entropy. That is, the system goes from a pure state initially, with entropy of zero, to a mixed state with nonzero entropy. 

Following Eq. (\ref{eq:infodeco}), we see that determining the mutual information between the system and the apparatus amounts to finding the entropy for the reduced state of either the system or the apparatus. In a general measurement scheme, these may depend on time, but when the system and apparatus states undergo \emph{independent} unitary evolution after the measurement process, we need only consider the states immediately after the measurement.\cite{Note13} In our case, for example, the particles passing through the slits evolve as free particles after the interaction, while the apparatus state remains unchanged. Hence, we may compute the mutual information from the state of the apparatus immediately after the measurement, without involving the more complicated time-dependent state of the particle.

\begin{figure}[htp]
  \begin{center}
    \includegraphics[width=0.9\columnwidth]{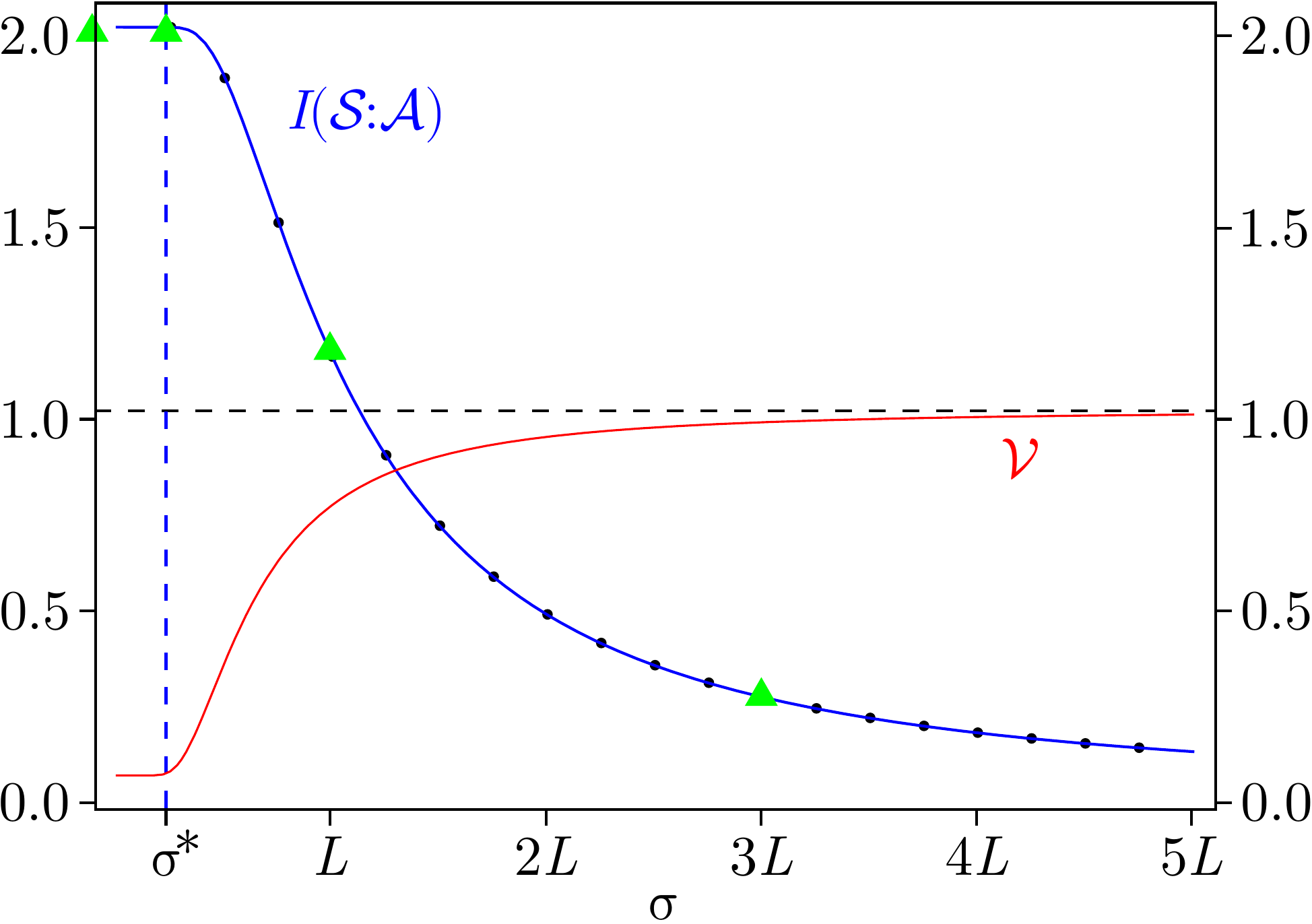}
    \caption{The dependence of mutual information \(I\left(\S:\A\right)\) and visibility \(\mathcal{V}\) on \(\sigma\). Extremely precise measurements, corresponding to small values of \(\sigma\), result in the apparatus acquiring significant information about the system while the interference is negligible. As \(\sigma\) increases, the measurement is less capable of distinguishing the particle's path, and the apparatus fails to decohere the system state. Hence, less information is transferred into the apparatus and the interference becomes significant. Visibility is calculated according to the exact expression in Eq. (\ref{eq:visbeta}) with \(L = 5\). The vertical dashed line at \(\sigma^\star \approx 3L/10\) indicates the approximate precision at which this transition begins to become apparent. Dots indicate values calculated numerically from the exact apparatus state, Eq. (\ref{eq:postapp}), for \(L = 5\). Triangles correspond to \(\sigma\) values considered in Fig. \ref{fig:measured_particle}.}
    \label{fig:MI_Vis}
  \end{center}
\end{figure}

We have previously already found the eigenvalues of the state \(\rho_\A \) after measurement has occurred, which are given in Eq.~(\ref{eq:appeigens}). These eigenvalues give the mutual information 
\[
  I\left(\S:\A\right) = 2H(\op{\rho}_\A) = 2 \left( -\lambda_+ \log_2 \lambda_+ -\lambda_- \log_2 \lambda_- \right).
\]
In the limit of no measurement, \(\sigma \to \infty\), the eigenvalues are \(1\) and \(0\), so the mutual information is zero: the apparatus stores no information about the state of the system. This limit is precisely that in which the standard interference pattern is observed. On the other hand, \(\sigma \to 0\) corresponds to the complete absence of interference and the eigenvalues monotonically approach \((1 \pm e^{-L^2})/2 \approx 1/2\), in which case the mutual information approaches \(I\left(\S:\A\right) = 2\). The dependence of the mutual information and visibility on the precision \(\sigma\), is shown in Fig. \ref{fig:MI_Vis}, demonstrating that as the information gained by the apparatus decreases, the visibility of the interference increases.

As is shown in Fig.~\ref{fig:MI_Vis}, both $\MI$ and $\mathcal{V}$ are initially flat as $\sigma$ increases from \(0\), as at small \(\sigma\) the measurement is extremely precise and the particle at the left and right slits can be effectively distinguished. Beyond a threshold value, \(\sigma = \sigma^\star \approx 3L/10\),\cite{Note14} the apparatus rapidly loses the ability to distinguish between the two paths, and interference begins to emerge.

Considering the above discussion, one notes that the \emph{quantum} mutual information indicates that, in the case of perfect measurement, we get \emph{two} bits of information, despite the fact that the information in which we are interested appears to be a simple binary statement regarding the particle's path, or one bit. This is a peculiarity of quantum information, corresponding to the existence of non-classical correlations (entanglement) between the apparatus and the system. If there is a third link in the von Neumann chain---e.g., an observer making measurements on \emph{the apparatus}---we will find that there is only one bit of information between the system and apparatus or between the system and observer. Indeed, that there are many links in the von Neumann chain, including not just the apparatus and observer, but also the large surrounding environment, e.g., photons, is why quantum correlations are so hard to detect.\cite{Zwolak2013} The presence of many such links is reflected in the redundant acquisition of information by the environment, which is the quantum Darwinian process responsible for the emergence of the classical, objective world.\cite{Zurek2014}. To see this, consider an observer \(\O\) that perfectly measures the apparatus immediately after the particle has been measured. In other words, if the observer is initially in the state \(\ket{0}_\O\), the observer and the apparatus evolve according to
\begin{align}
  \ket{L, 0}_{\A\O} &\mapsto \ket{L, L}_{\A\O},\nonumber\\
  \ket{R, 0}_{\A\O} &\mapsto \ket{R, R}_{\A\O}.
\end{align}
Then we should replace Eq. (\ref{eq:unitarymeasure}) with
\begin{equation}
    \ket{\Psi,0,0}_{\S\A\O} \mapsto \frac{1}{2}\ket{\Psi^L, L,L}_{\S\A\O} + \frac{1}{2}\ket{\Psi^R, R,R}_{\S\A\O},\label{eq:GlobalState}
\end{equation}
where for a perfect measurement the system states \(\ket{\Psi^L}\) and \(\ket{\Psi^R}\) would be orthogonal.

If we now compute the partial trace over the apparatus as done in Eq. (\ref{eq:poststate}), we find that the joint system-\emph{observer} state is
\begin{equation}
\op{\rho}_{\S\O} = \frac{1}{2}\ketbra[\S\O]{\Psi^L, L}{\Psi^L, L} + \frac{1}{2}\ketbra[\S\O]{\Psi^R, R}{\Psi^R, R}.
\end{equation}
A second partial trace over the observer will recover Eq. (\ref{eq:poststate}), showing that the entropy of the particle is unaffected by the observer's measurement of the apparatus. The entropy of the observer, though, is \(H(\rho_\O) = 1\). In contrast to the previous discussion, the joint state \(\rho_{\S\O}\) is not pure, so its entropy does not vanish. Rather, the orthogonality of \(\ket{\Psi^L, L}_{\S\O}\) and \(\ket{\Psi^R, R}_{\S\O}\), due to the presence of the left/right record in the observer's state, indicate that this, too, has \(H(\rho_{\S\O}) = 1\). Hence, the mutual information is
\begin{align}
  I\left(\S:\O\right) &= H(\op{\rho}_{\S}) + H(\op{\rho}_{\O}) - H(\op{\rho}_{\S\O})\\
  &= H(\op{\rho}_\S) + 1 - 1 = H(\op{\rho}_\S), \label{eq:MIwO}
\end{align}
indicating that the observer acquires an amount of information equal to what is available about the path of the particle, \( H(\op{\rho}_\S) \). If path information is present, or the measurement precision is \(\sigma \to 0 \) yielding \( H(\op{\rho}_\S) \to 1 \), then the observer will acquire 1 bit of information. If not, \( H(\op{\rho}_\S) \approx 0 \), the observer will learn nothing about the path of the particle and interference will be observed at the screen. When the observer is present and the global state is Eq. (\ref{eq:GlobalState}), the apparatus will also have mutual information given by Eq. (\ref{eq:MIwO}), as entanglement with $\S$ is ``locked up'' in joint correlations between $\S$ and $\A \O$.

Finally, we note that in the large \(L\) approximation, Eqs. (\ref{eq:appeigens}) and (\ref{eq:visbeta}) together allow us to write
\begin{align}
  I\left(\S:\O\right) &= H_{bin}\left(\frac{1 + \mathcal{V}}{2}\right) .
\end{align}
This explicitly connects the information gain by the observer with the loss of visibility of the interference fringes through the binary entropy
\begin{equation}
  H_{bin}(x) = -x\log_2{x} - (1 - x)\log_2{(1 - x)},
\end{equation}
which characterizes the uncertainty regarding the outcome of a classical event that could result in one of two outcomes with probabilities \(x\) and \(1 - x\), respectively.
When the fringes are readily apparent \(\mathcal{V} \approx 1\) and the information acquired by the observer (or apparatus) is \(I\left(\S:\O\right) \approx 0\). On the other extreme, when the fringes are not visible \(\mathcal{V} \approx 0\) and the information gain is \(I\left(\S:\O\right) \approx 1\). Note also that in the intermediate regimes one can have quite high visibilities even for \(I\left(\S:\O\right)\) near \(1\), but for \(\sigma \approx L/2\) there are still visible interference fringes despite gaining nearly complete information about the system's path. This fact was also noted in the case of interference of photons.\cite{Wootters1979}

\section{Conclusion}
By considering a specific model of double-slit interference, we have shown how the precision with which one determines the path of particles passing through the slit is directly correlated with the loss of observed interference effects in the subsequent evolution of the particles. In particular, we have shown how the absence of interference can be attributed to an apparatus gaining maximal information gaining maximal information, while a measurement that acquires no information about the system has no effect on the interference (note that the apparatus may be a physical device, an observer, or the environment) .

It bears mentioning that while we have focused exclusively on how the act of measurement can cause a loss of interference, this is by no means the only reason interference may not be observed. Another cause for interference loss is \emph{dephasing}, which occurs when, for example, the relative phase between the two Gaussians in the superposition varies from trial to trial. In this case, the absence of interference is a statistical result arising from the oscillation term acquiring  a different phase in each trial, which causes the probability density to be shifted. If the phase is Gaussian distributed with a width \(\gamma\) (and mean 0), then the expected probability density for a free particle (\(\beta_\sigma = 1\)) becomes
\begin{equation}
  \expect{P} \approx \Gamma_{xt}\left[\cosh{\left(\frac{2xL}{1 + t^2}\right)} +  {e}^{-\frac{1}{2}\gamma^2}\cos{\left(\frac{2xLt}{1 + t^2}\right)}\right],
\end{equation}
when \(L \gg 1\) (this calculation is similar to that in Sect.~\ref{sec:Free_Interference}).
Hence, a sharply-peaked distribution of phases will exhibit interference that becomes washed out as the distribution widens. While this ``dephasing'' process produces a similar experimental outcome (namely, the loss of interference), it is important to note that the physical process is quite different than that of measurement-induced decoherence.\cite{Schlosshauer2010} In particular, decoherence removes interference from the wavefunction for every trial, whereas the loss of interference due to dephasing is found only as a result of averaging over many different trials.

The model we have examined serves as a concrete example of the relationship between information and interference  in quantum systems. It is approachable by students in the latter portion of introductory courses (such as the second or third course in a year-long sequence), including those at the upper-undergraduate level in many programs. It can serve as a basis for homework problems and projects by considering, for example, small \(L\) rather than large \(L\) approximations, different amplitudes in the initial superposition, more general measurement schemes, application of numerical and approximation techniques, or a double-well preparation of the initial state (e.g., using two delta function potentials), among other things. It will thus provide a link between the usual conceptual discussion surrounding the double slit experiment and actual calculations, as well as bring a modern account of measurement into the classroom.


\begin{thebibliography}{99}

\bibitem{Young1802} T. Young, ``The Bakerian Lecture: On the Theory of Light and Colours,'' Philos. Trans. Roy. Soc. London Ser. A \textbf{92}, 12--48 (1802)

\bibitem{Joensson1961} C. J\"onsson, ``Elektroneninterferenzen an mehreren künstlich hergestellten Feinspalten,'' Z. Phys. \textbf{161}, 454--474 (1961)

\bibitem{Tonomura1989} A. Tonomura \textit{et al}, ``Demonstration of single‐electron buildup of an interference pattern,'' Amer. J. Phys. \textbf{57}, 117--120 (1989)

\bibitem{Arndt1999} M. Arndt \textit{et al}, ``Wave-particle duality of C60~{mol}ecules,'' Nature \textbf{401},\ \bibinfo {pages} {680--682} (1999)

\bibitem{Eibenberger2013} S. Eibenberger \textit{et al}, ``Matter-wave interference of particles selected from a molecular library with masses exceeding 10000 amu,'' Phys. Chem. Chem. Phys. \textbf{15}, 14696--14700 (2013)
  
\bibitem{Arndt2014} M. Arndt and K. Hornberger, ``Testing the limits of quantum mechanical superpositions,'' Nat. Phys. \textbf{10}, 271--277 (2014)

\bibitem{Beugnon2006} J. Beugnon \textit{et al}, ``Quantum interference between two single photons emitted by independently trapped atoms,'' Nature \textbf{440}, 779--782 (2006)

\bibitem{Maunz2007} P. Maunz \textit{et al}, ``Quantum interference of photon pairs from two remote trapped atomic ions,'' Nat. Phys. \textbf{3}, 538--541 (2007)

\bibitem{Andrews1997} M. R. Andrews \textit{et al}, ``Observation of Interference Between Two Bose Condensates,'' Science \textbf{275}, 637--641 (1997)

\bibitem{Dziarmaga2012} J. Dziarmaga, W. H. Zurek, and M. Zwolak, ``Non-local quantum superpositions of topological defects,'' Nat. Phys. \textbf{8}, 49--53 (2012)

\bibitem{Feynman1963c} R. P. Feynman, R. B. Leighton, and M. Sands, 
\textit{The Feynman Lectures on Physics, Vol.\ 3} (Addison-Wesley, Boston, 1964).

\bibitem{Wootters1979} W. K. Wootters and W. H. Zurek, ``Complementarity in the double-slit experiment: Quantum nonseparability and a quantitative statement of Bohr's principle,'' Phys. Rev. D \textbf{19}, 473--484 (1979)

\bibitem{Raymer1992} M. G. Raymer and S. Yang, ``Information and Complementarity in a Proposed Which-path Experiment Using Photons,'' J. Modern Opt. \textbf {39}, 1221--1231 (1992)

\bibitem{Note1} See Chapter 2.6.2 of Ref. \onlinecite{Schlosshauer2010}.

\bibitem{Ollivier2004} H. Ollivier, D. Poulin, and W. H. Zurek, ``Objective Properties from Subjective Quantum States: Environment as a Witness,'' Phys. Rev. Lett. \textbf{93}, 220401 (2004)

\bibitem{Ollivier2005} H. Ollivier, D. Poulin, and W. H. Zurek, ``Environment as a witness: Selective proliferation of information and emergence of objectivity in a quantum universe,'' Phys. Rev. A \textbf{72}, 042113 (2005)

\bibitem{Blume-Kohout2006} R. Blume-Kohout and W. H. Zurek, ``Quantum Darwinism: Entanglement, branches, and the emergent classicality of redundantly stored quantum information,'' Phys. Rev. A \textbf{73}, 062310 (2006)

\bibitem{Zurek2009} W. H. Zurek, ``Quantum Darwinism,'' Nat. Phys. \textbf{5}, 181--188 (2009)

\bibitem{Zwolak2013} M. Zwolak, W. H. Zurek, ``Complementarity of quantum discord and classically accessible information,'' Sci. Rep. \textbf{3}, 1729 (2013)

\bibitem{Zwolak2014} M. Zwolak, C. J. Riedel, W. H. Zurek, ``Amplification, Redundancy, and Quantum Chernoff Information,'' Phys. Rev. Lett. \textbf{112}, 140406 (2014)

\bibitem{Zurek2014} W. H. Zurek, ``Quantum Darwinism, classical reality, and the randomness of quantum jumps,'' Phys. Today \textbf{67} (10), 44--50 (2014)

\bibitem{Zurek1981} W. H. Zurek, ``Pointer basis of quantum apparatus: Into what mixture does the wave packet collapse?,'' Phys. Rev. D \textbf{24}, 1516--1525 (1981)

\bibitem{Zurek1982} W. H. Zurek, ``Environment-induced superselection rules,'' Phys. Rev. D \textbf{26}, 1862--1880 (1982)
  
\bibitem{Zurek1991} W. H. Zurek, ``Decoherence and the Transition from Quantum to Classical,'' Phys. Today \textbf{44} (10), 36--44 (1991)

\bibitem{Albrecht1992} A. Albrecht, ``Investigating decoherence in a simple system,'' Phys. Rev. D \textbf{46}, 5504--5520 (1992)

\bibitem{Tegmark1994} M. Tegmark and H. S. Shapiro, ``Decoherence produces coherent states: An explicit proof for harmonic chains,'' Phys. Rev. E \textbf{50}, 2538--2547 (1994)

\bibitem{Schumacher1996} B. Schumacher, M. Westmoreland, and W. K. Wootters, ``Limitation on the Amount of Accessible Information in a Quantum Channel,'' Phys. Rev. Lett. \textbf{76}, 3452--3455 (1996)

\bibitem{Zurek2003} W. H. Zurek, ``Decoherence, einselection, and the quantum origins of the classical,'' Rev. Modern Phys. \textbf{75}, 715--775 (2003)

\bibitem{Joos2003} E. Joos \textit{et al}, \textit{Decoherence and the Appearance of a Classical World in Quantum Theory}, (Springer-Verlag, Berlin, 2003)

\bibitem{Nielsen2010} M. A. Nielsen and I. L. Chuang, \textit{Quantum Computation and Quantum Information}, (Cambridge University Press, Cambridge, 2010)

\bibitem{Schlosshauer2010} M. Schlosshauer, \textit{Decoherence and
  the Quantum-to-Classical Transition}, (Springer-Verlag, Berlin, 2010)

\bibitem{Note2} See Chapter 2.4 of Ref. \onlinecite{Nielsen2010}.

\bibitem{Shankar1994} R. Shankar, \textit{Principles of Quantum Mechanics}, (Plenum Press, New York, 1994), pp 659--660

\bibitem{Note4} See Page 105 of Ref. \onlinecite{Nielsen2010}.

\bibitem{Note5} See Chapter 11 of Ref. \onlinecite{Nielsen2010}.

\bibitem{Note6} The exact wavefunction emerging from the slits in an experimental set-up will be complicated, depending on the exact slit shape geometry, the incoming state, particle type, etc. A superposition of Gaussian wavepackets, which we consider, leads to a clean presentation of interference and measurement. A square wavepacket, for \(\psi = \Theta(x \pm L + \Delta) - \Theta(x \pm L - \Delta)\), with \(\Theta\) the Heaviside step-function at each slit, which matches the ideal geometry of the slit, is also tractable, and can give a suitable extension of this work for class projects. The actual wavefunction, however, will be more complicated. Indeed, a reasonable approximate form is \(\psi = \exp{\left(-1/[\alpha((\Delta/2)^2-(x\mp L)^2)]\right)}\) (and zero for \(x >\pm L+\Delta/2\) and \(x < \pm L - \Delta/2\)) at each slit of width \(\Delta\). This function, while only approximate, is already non-analytic, but can both match the geometry and have smooth boundaries.

\bibitem{Neumann1932} J. von Neumann, \textit{Mathematische Grundlagen
  der Quantenmechanik} (Verlag von Julius Springer,
  Berlin, 1932)

\bibitem{Note7} Recall that the partial trace is defined by
\[
\tr_\A(\rho_{\S\A}) = \sum_k{\prescript{}{\A}{\bra{k}}\rho_{\S\A}\ket{k}_\A},
\]
where \(\{\ket{s_i}_\A\}\) is any basis for the apparatus Hilbert space, resulting in an operator that acts only on the system. For example, one has
\[
\tr_\A{\left(\ketbra[\S\A]{\M^L\Psi, L}{\M^R\Psi, R}\right)} = 0,
\]
as the left hand side is equal to $\M^L\ketbra[\S]{\Psi}{\Psi}\adj{\M^R}\tr{\left(\ketbra[\A]{L}{R}\right)}$.
Note that the states \unexpanded{\(\ket{L}_\A\)} and \unexpanded{\(\ket{R}_\A\)} in Eq. (\ref{eq:poststate}) do not constitute a complete basis for our apparatus state, but they do give the only nonzero terms in the trace here.

\bibitem{Caves1987} C. M. Caves, and G. J. Milburn, ``Quantum-mechanical model for continuous position measurements,'' Phys. Rev. A \textbf{36}, 5543--5555 (1987)

\bibitem{Dowker1992} H. F. Dowker and J. J. Halliwell, ``Quantum mechanics of history: The decoherence functional in quantum mechanics,'' Phys. Rev. D \textbf{46}, 1580--1609 (1992)

\bibitem{Note8} The first-order approximation about \(x = -L\) is given in Eq. (\ref{eq:linear_m}) with the replacement \(L\mapsto-L\). The linear coefficient can easily be seen to be bounded graphically, but in this case the denominator includes a factor of \(m_\sigma(-L)\), which cannot be bounded below by a positive constant. A numerical calculation provides a bound of approximately \(0.379/L\) on the actual coefficient, and there are many functions that can be used to find bounds analytically. For example, inserting the bound
\[
  \Erfc{\left(x/\sqrt{2}\right)}>\sqrt{2/\pi}(x/(x^2 + 1))e^{-x^2/2}
\]
into the linear coefficient and maximizing yields a bound slightly lower than the \(0.4/L\) used in the main text.

\bibitem{Note9} See supplementary material at \protect\url{http://dx.doi.org/10.1119/1.4943585} for a calculation providing dynamic visualization of the distribution for arbitrary parameters.

\bibitem{Note10} See Chapter 2.2.6 of Ref. \onlinecite{Nielsen2010}.

\bibitem{Note11} We note that this statement assumes the system state is initially pure.

\bibitem{Schmidt1908} E. Schmidt, ``Zur Theorie der linearen und nichtlinearen Integralgleichungen,'' Math. Ann. \textbf{65}, 370--399 (1908)

\bibitem{Ekert1995} A. Ekert and P. L. Knight, ``Entangled quantum systems and the Schmidt decomposition,'' Amer. J. Phys. \textbf{63}, 415--423 (1995)

\bibitem{Note12} See Chapter 2.5 of Ref. \onlinecite{Nielsen2010}.

\bibitem{Note13} Students reading this might find it enlightening to perform the computation showing that the entropies remain unchanged by independent evolution of the system and apparatus

\bibitem{Note14} The estimate for \(\sigma^\star\) is found by observing that small values of \(\sigma\) correspond to large values in the argument of \(\Erfc\). Hence, the small-\(\sigma\) case can be analyzed using an asymptotic expansion, leading to
\[
\beta_\sigma \approx \sqrt{\frac{2\sigma}{L}\sqrt{\frac{2}{\pi}}}\exp{\left(-L^2/4\sigma^2\right)}.
\]
Further expanding this expression as a Taylor series about \(\sigma = L/2\) and locating the \(x\)-intercept then gives
\[
  \beta_{\sigma^\star} \approx \left(\frac{2}{\pi}\right)^{1/4}{e}^{-1}\left[1 + \frac{5}{L} \left(\sigma^\star - \frac{L}{2}\right)\right] = 0,
\]
so we find
\[
  \sigma^\star \approx 3L/10,
\]
which is the value stated in the main text.
\end{thebibliography}
\end{document}